\documentclass{PoS}

\usepackage{amsmath,graphicx}
\newcommand{\feynslash}[1]{/\hspace*{-1.9mm} #1}
\newcommand{\vv }{\mathrm{v}}

\title{Heavy-quark symmetries in the light of nonperturbative QCD approaches}

\ShortTitle{Heavy-quark symmetries in the light of nonperturbative QCD approaches }

\author{\speaker{Bruno El-Bennich} \\
         Laboratorio de F\'isica Te\'orica e Computa\c{c}\~ao Cient\'ifica, Universidade Cruzeiro do Sul, 01506-000, S\~ao Paulo, Brazil \\
         Instituto de F\'isica Te\'orica, Universidade Estadual Paulista, 01140-070, S\~ao Paulo, Brazil \\
         E-mail: \email{bruno.bennich@cruzeirodosul.edu.br}}

\author{Craig~D.~Roberts\\
        Physics Division, Argonne National Laboratory, Argonne, Illinois 60439, USA \\    
        Department of Physics, Illinois Institute of Technology, Chicago, Illinois 60616-3793, USA \\
        E-mail: \email{cdroberts@anl.gov}}

\author{Mikhail~A.~Ivanov \\
        Bogoliubov Laboratory of Theoretical Physics, Joint Institute for Nuclear Research, 141980 Dubna, Russia \\
        E-mail: \email{ivanovm@theor.jinr.ru}}

\abstract{We critically review the validity of heavy-quark spin and flavor symmetries in heavy-light decay constants, form factors and effective couplings obtained
                 within a nonperturbative framework, the ingredients of which are all motivated by Dyson-Schwinger equations studies of QCD. Along the way, we make new 
                 predictions for two effective nonphysical couplings: $g_{D_sDK} = 24.1^{+2.5}_{-1.6}$ and $g_{B_sBK} = 33.3_{-3.7}^{+4.0}$.}

\FullConference{International Workshop on QCD Green's Functions, Confinement and Phenomenology,\\
		           September 05--09, 2011\\
		           Trento, Italy}

\begin{document}

\section{Introduction: heavy-quark symmetries}
\label{sec1}

It is common knowledge that the lightest quarks with the masses $m_u$ and $m_d$ are small compared to the nonperturbative scale $\Lambda_\mathrm{QCD}$ 
dynamically generated in Quantum Chromodynamics (QCD), which leads to an $SU(2)_L\times SU(2)_R$ chiral symmetry. This is frequently extended to
include the strange quark which forms an $SU(3)_L\times SU(3)_R$ symmetry. On the other hand, for heavy quarks with masses $m_Q \gg \Lambda_\mathrm{QCD}$ 
it is also common practice to take the limit $m_Q\to \infty$ as a good approximation~\cite{Manohar:2000dt}. Heavy-spin and -flavor symmetries are not manifest in 
the QCD Lagrangian even in the limit $m_Q \to \infty$. These symmetries emerge in effective field theories for QCD, for example heavy quark effective theory (HQET) 
which describes the dynamics of mesons containing a single heavy quark.

The interactions of heavy and light quarks in $Q\bar q$ mesons are governed by nonperturbative dynamics and are of the order $\Lambda_\mathrm{QCD}$,
which incidentally is also the heavy mesons's size scale. Thus, in the $m_Q \to \infty$ limit the velocity of the heavy quark is practically unchanged by the 
interactions since $\Delta \vv =\Delta p/m_Q$. In effect, the four-momentum $p_\mu=m_Q \vv_\mu$ of an on-shell quark is altered by the interaction to that of an 
off-shell quark with momentum $p_\mu=m_Q \vv_\mu +k_\mu$, where $k \sim \Lambda_\mathrm{QCD}$ is the residual momentum. In the heavy-quark limit, the
usual quark propagator becomes,
\begin{equation}
   S(p) \, = \, i \frac{\feynslash \! p + m_Q}{p^2-m_Q^2}\  \stackrel{m_Q\to \infty}{\longrightarrow}\ i\, \frac{1+\feynslash\vv }{2\, \vv \cdot k}\ + \ \mathcal{O}\left (\frac{k}{m_Q}\right )\ ,
 \label{mqpropagator}
\end{equation}
where $\vv _\mu$ is a time-like unit vector, $\vv^2 = 1$, for instance $\vv _\mu=(1,\vec 0)$ in the heavy quark's rest frame. An effective theory for heavy fields is obtained from 
the original Lagrangian by separating the large and small components of the Dirac spinor fields with the projectors,
\begin{equation}
  h_\vv (x) = e^{im_Q \vv \cdot x}\, \frac{1+\feynslash \vv }{2}\, Q(x)\ , \qquad H_\vv (x) = e^{im_Q \vv \cdot x}\, \frac{1- \feynslash \vv }{2}\,  Q(x) \ ,
\end{equation}
where the exponential prefactor serves to subtract $m_Q \vv_\mu$ from the heavy quark momentum. The new fields thus carry residual momenta $k\sim\Lambda_\mathrm{QCD}$
and satisfy the relations: 
\begin{equation}
\feynslash  \vv  h_\vv (x) = h_\vv (x) \qquad \feynslash  \vv  H_\vv (x) = -H_ \vv (x) \ . 
\label{vrelations}
\end{equation}
Inserting these effective two-component fields into the part of the QCD Lagrangian which involves the heavy-quark fields and using the relations (\ref{vrelations}) 
and $\bar h_\vv \, \feynslash \vv \, H_\vv = 0$ leads to a decomposition into massless and massive terms,
\begin{equation}
 \mathcal{L}_Q = \bar Q (\feynslash\! D - m_Q) Q \ = \ \underbrace{\bar h_\vv \, i\,  \vv \cdot D\, h_\vv }_{\textrm{\scriptsize massless mode}} \ +\  
         \underbrace{\bar H_\vv \,(- i\, \vv \cdot D - 2m_Q)H_v}_{\textrm{\scriptsize massive mode}} \ + \    
         \underbrace{\bar h_\vv \, i\, \feynslash\! \vec D\, H_\vv  + \bar H_\vv \, i\, \feynslash\! \vec D\, h_\vv }_{\textrm{\scriptsize interaction terms}} \ ,
\label{lagrangian} 
\end{equation}
where $m_Q$ is the renormalized heavy-quark current mass. Note that the Feynman rules for 
the massless mode yield the propagator of Eq.~(\ref{mqpropagator}) in the $m_Q\to \infty$ limit. 

To derive the effective Lagrangian one eliminates the massive modes by solving the equation of motion for $H_\vv (x)$, {\em i.e.\/} 
$(i\, \vv \cdot D + 2m_Q) H_\vv  = i\, \feynslash\! \vec D\, h_\vv $, where on the right-hand-side we used the definition $(0\,;\, \vec D)\equiv D^\mu-\vv\cdot D \,\vv ^\mu$. 
Moreover, since the momenta $k$ of the fields $h_\vv (x)$ are much smaller than $m_Q$, an expansion in local derivative operators is justified:
\begin{equation}
  H_\vv  = \frac{1}{i\, \vv \cdot D+ 2m_Q}\, i\ \feynslash\!\vec D\, h_\vv  \ = \  \frac{1}{2m_Q}\,\sum_{n=0}^\infty \left (- \frac{i\, \vv \cdot D}{2 m_Q} \right )^{\!\!n}  i\ \feynslash\!\vec D\, h_\vv  \ .
\end{equation}
Replacing $H_\vv (x)$ in the QCD Lagrangian (\ref{lagrangian}) by this expansion, we arrive at the HQET Lagrangian of a heavy quark $h_\vv (x)$ interacting with
soft gluons~\cite{Manohar:2000dt,Neubert:1993mb},
\begin{equation}
  \mathcal{L}_{\mathrm{\scriptsize HQET}} \ = \  \bar h_\vv \, i \vv \cdot D\, h_\vv \  + \ \frac{1}{2m_Q}  \Big [ \bar h_\vv \, (i\,{\vec D})^2\, h_\vv 
    + c(\zeta) \frac{g}{2}\, \bar h_\vv \, \sigma_{\mu\nu} g^{\mu\nu} h_\vv   \Big ]  \ + \ \ ... \  \ ,
 \label{lagrangeHQET}
\end{equation}
an expansion in powers of $\alpha_s$ and $m_Q^{-1}$. Here, $c(\zeta)$ denotes higher-order operator corrections in $\alpha_s(\zeta) =g^2/4\pi$ where the matching 
condition at tree level implies $c(m_Q) =1 + \mathcal{O}[\alpha_s(m_Q^2)]$ and $\zeta$ is the renormalization point. Plainly, in the heavy-quark limit only the first term 
survives in Eq.~(\ref{lagrangeHQET}). Since there is no reference to the heavy mass in this term and $h_\vv(x)$ is invariant under spin rotation, the leading order 
Lagrangian exhibits the full $U(2N_h)$ spin-flavor symmetry ($N_h$: heavy flavor number). This symmetry is broken by nonperturbative $\mathcal{O}(m_Q^{-1})$ 
and $\alpha_s$ corrections of order  $\Lambda_\mathrm{QCD}/m_Q$: the second term in Eq.~(\ref{lagrangeHQET}) is the heavy quark kinetic energy ${\vec p\,}^2\!/2m_Q$ 
in a nonrelativistic constituent quark model and breaks the heavy quark flavor symmetry; the third term is the heavy quark's chromomagnetic interaction with the 
gluon field and breaks both heavy quark spin and flavor symmetries.

These symmetries are also broken at the perturbative level and in general the hard gluon exchanges renormalize coefficients that multiply effective operators.
This was already seen in Eq.~(\ref{lagrangeHQET}) for the chromomagnetic term, where the renormalization point dependence of $c(\zeta)$ must be cancelled
by that of the magnetic moment operator. However, in practical applications the operators give rise to form factors whose exact $\zeta$ dependence gets lost
depending on the nonperturbative approach to their calculation. NB: the leading order and kinetic energy terms are not renormalized by gluon corrections besides 
the usual quark wavefunction renormalization; this is due to reparameterization ($\equiv$~Lorentz) invariance.

To summarize, heavy quark spin-flavor symmetries emerge as a property of the leading term in an $m_Q^{-1}$ expansion of the heavy-quark sector in the QCD
Lagrangian. In the $m_Q\to \infty$ limit, the heavy-quark velocity becomes a conserved quantity and the momentum exchange with surrounding light degrees of 
freedom is predominantly soft. Since the heavy-quark spin decouples in this limit, light quarks are blind to it. In essence, they do not experience any different 
interactions with a much heavier quark in a pseudoscalar or vector meson. Practically, this implies certain symmetries between hadronic observables. For example,
the decay constants of the heavy mesons should {\em sensu stricto\/} read: $f_D=f_{D^*}$  and $f_B=f_{B^*}$. Whereas the last relation may be sensible due to the 
magnitude of $\Lambda_\mathrm{QCD}/m_b$ corrections, the first relation is questionable --- the charm-quark mass is known to lie in an uncomfortable mass region, 
neither really heavy nor light. Nonetheless, HQET is frequently employed in calculations of matrix elements which involve charmed $Q\bar q$ states. In Section~\ref{sec4} 
we discuss from an exclusively nonperturbative perspective whether this expansion is justified and to which degree it is verified.

\section{Flavored Dyson-Schwinger Equations}
\label{sec2}

Poincar\'e-covariant and symmetry-preserving models built upon robust predictions of QCD's Dyson-Schwinger equations (DSEs) provide a well-grounded framework 
within which heavy $Q\bar q$ bound states can be examined. A large set of heavy-meson observables \cite{Ivanov:1997yg,Ivanov:1997iu,Ivanov:1998ms,Ivanov:2007cw,
ElBennich:2010ha,ElBennich:2011py} has been studied in this nonperturbative modeling approach. They possess the feature that quark propagation is described by dressed 
Schwinger functions, which has a material impact on light-quark characteristics and hadron phenomenology~\cite{Roberts:1994hh,Roberts2000,Roberts:2007ji,
Roberts:2007jh,Chang:2009ae,Cloet:2008re,Chang:2012rk}. It is also in contrast with lattice-regularized QCD simulations which treat the $b$-quark as static
\cite{Becirevic:2009xp,Becirevic:2009yb,Albertus:2010nm,Dimopoulos:2011gx} and the $c$-quark as a propagating mode but its dynamics is quenched~\cite{Becirevic:2012ti}.

We here briefly summarize the dressed Schwinger functions ($\equiv$ Euclidean Green function) obtained from DSEs solutions in QCD. A general review of the DSEs can 
be found in Refs.~\cite{Roberts:2007ji,Roberts:1994dr} and their applications have recently been surveyed for hadron physics in general~\cite{Roberts:2007jh,Bashir:2012fs}
and heavy mesons in particular~\cite{Bashir:2012fs,ElBennich:2009vx}. The (anti)quarks' dressing in $q\bar q$ and $Q\bar q$ bound states is described for a given 
quark flavor by the DSE,\footnote{\, Henceforth, we employ Euclidean metric in our notation: $\{\gamma_\mu,\gamma_\nu\} = 2\delta_{\mu\nu}$; $\gamma_\mu^\dagger = \gamma_\mu$; 
$\gamma_5= \gamma_4\gamma_1\gamma_2\gamma_3$, tr$[\gamma_4\gamma_\mu\gamma_\nu\gamma_\rho\gamma_\sigma]=-4 \epsilon_{\mu\nu\rho\sigma}$; 
$\sigma_{\mu\nu}=(i/2)[\gamma_\mu,\gamma_\nu]$; $a \cdot b = \sum_{i=1}^4 a_i b_i$; and $P_\mu$ timelike $\Rightarrow$ $P^2<0$.}
\begin{equation}
S^{-1}(p)  =   Z_2 (i\, \feynslash \! p + m^{\mathrm{bm}}) + \Sigma (p^2) \ ,
\label{DSEquark}
\end{equation}
where the dressed quark self energy is obtained from
\begin{equation}
 \Sigma (p^2) = Z_1\, g^2\! \!\int^\Lambda_q \!\!\ D^{\mu\nu}Ê(p-q) \frac{\lambda^a}{2} \gamma_\mu S(q) \Gamma^a_\nu (q,p) ,
\end{equation}
and $\int_q^\Lambda\equiv \int^\Lambda d^4q/(2\pi)^4$ represents a Poincar\'e invariant regularization of the integral with the regularization mass scale $\Lambda$. 
The current quark bare mass  $m^{\mathrm{bm}}$ receives corrections from the self energy $\Sigma (p^2)$ in which the integral is over the dressed gluon propagator, 
$D_{\mu\nu}(q)$, the dressed quark-gluon vertex, $\Gamma^a_\nu (q,p)$, and $\lambda^a$ are the usual $SU(3)$ color matrices. The solution to the gap equation~(\ref{DSEquark}) 
are of the general form:
\begin{eqnarray}
  S(p)&  = & -i\, \feynslash \! p \ \sigma_V (p^2) + \sigma_S(p^2) =  \left [ i\, \feynslash \! p \  A(p^2) + B(p^2) \right ]^{-1}  .
  \label{sigmaSV}
\end{eqnarray}

The renormalization constants for the quark-gluon vertex, $Z_1(\zeta,\Lambda^2)$, and quark-wave function, $Z_2(\zeta,\Lambda^2)$, depend on
the renormalization point, $\zeta$, the regularization scale, $\Lambda$, and the gauge parameter, whereas the mass function $M(p^2) = B(p^2)/A(p^2)$ 
is independent of $\zeta$. Since QCD is asymptotically free, it is useful to impose the  renormalization condition,
\begin{equation}
  S^{-1}(p)|_{p^2=\zeta^2}  = i\, \feynslash \! p + m(\zeta^2) \ ,
\end{equation}
at large space-like $\zeta^2$  where $m(\zeta^2)$ is the renormalized running quark mass, so that for $\zeta^2\gg \Lambda_\mathrm{QCD}^2$ quantitative 
matching with pQCD results can be made.

Infrared dressing of light quarks has a {\em deep\/} impact on hadron physics~\cite{Roberts:1994hh,Roberts2000,Roberts:2007ji,Roberts:2007jh,Chang:2009ae,
Cloet:2008re,Chang:2012rk}: the quark-wave function renormalization, $Z(p^2) = 1/A(p^2)$, is suppressed while the dressed quark-mass function, $M(p^2) = B(p^2)/A(p^2)$, 
is enhanced in the infrared. This emergent phenomenon is known as dynamical chiral symmetry breaking (DCSB), a feature of QCD which is a prerequisite for 
a constituent quark mass scale. Both, numerical solutions of the quark DSE and simulations of lattice-regularized QCD~\cite{Zhang:2004gv}, predict this behavior 
of $M(p^2)$ and pointwise  agreement between DSE and lattice results has been explored in Refs.~\cite{Bhagwat:2003vw,Bhagwat:2004}. 

\begin{figure}[t]
\begin{center}
\includegraphics*[scale=0.5]{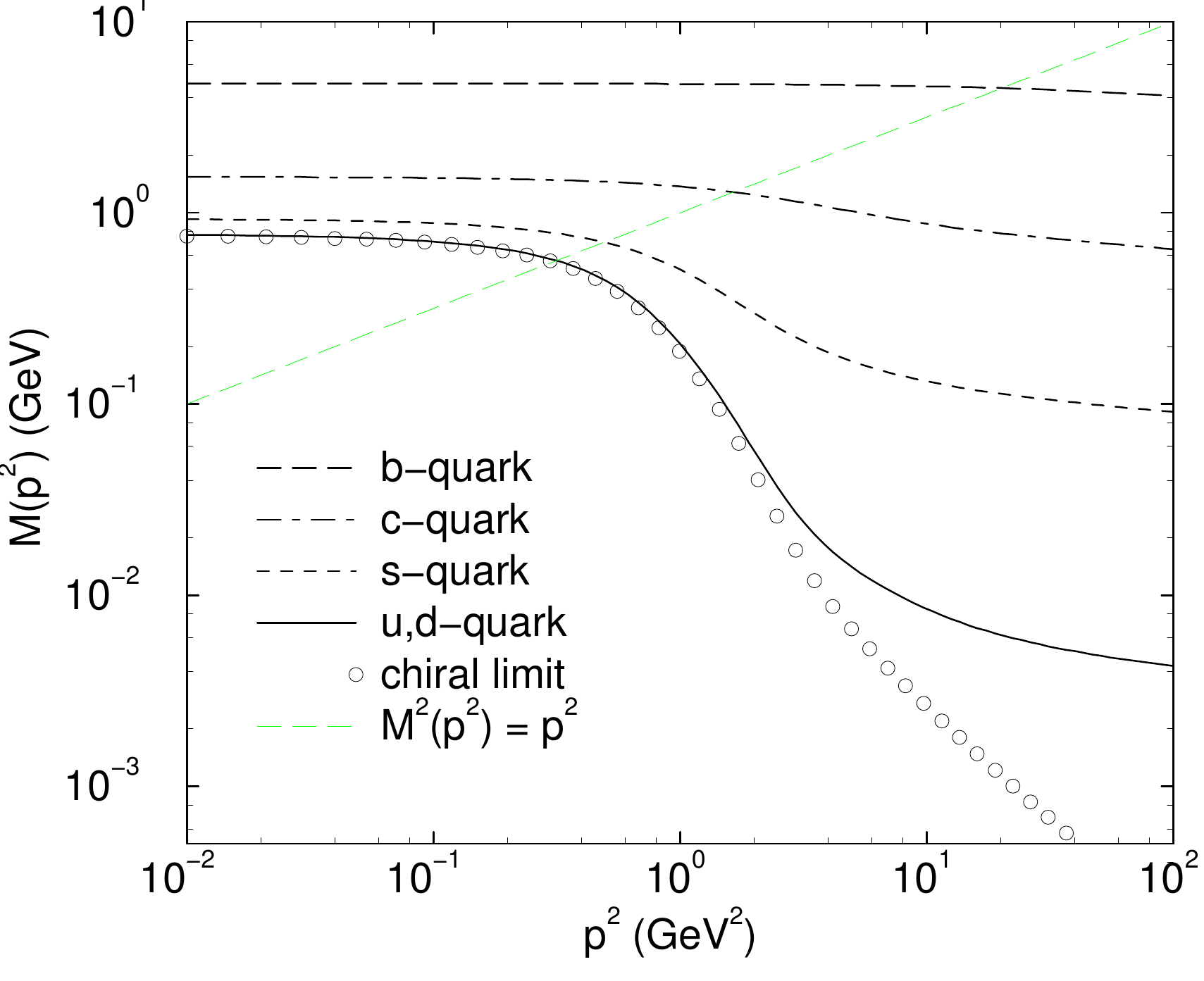}
\caption{Dynamical quark mass function generated in rainbow-ladder truncation for the flavours $q_f=u,d,s,c,b$ and in the chiral limit. 
See, for example, Refs.~\cite{Ivanov:1998ms,Roberts:2007jh} for details.}
\label{flavourmassfunc}
\end{center}
\end{figure}

Whereas the impact of gluon dressing is striking for light quarks, its effect on the heavy quarks is marginal. For the frequently used rainbow-ladder truncation 
of the gap equation~(\ref{DSEquark}) this can be appreciated, for instance, in Figure~\ref{flavourmassfunc}: for light quarks, mass can be generated from nothing, 
{\em i.e.\/}, the Higgs mechanism is irrelevant to their acquiring of a constituent-like mass. The effect of dressing the $c$-quark is modest and barely noticeable
for the $b$-quark, whose large current-quark mass almost entirely suppresses momentum-dependent dressing. Thus, $M_b(p^2)$ is nearly constant on a 
large domain, which is also true to a lesser extent for the charm quark.

In order to quantify this effect of DCSB, one can define the renormalization-point invariant ratio $\zeta_f := \sigma_f/M_f^E$, where $ \sigma_f$ is a 
constituent-quark $\sigma$-term~\cite{Holl:2005st}, 
\begin{equation}
 \sigma_f := m_f(\zeta ) \frac{\partial M_f^E}{\partial m_f(\zeta ) } \ ,
\end{equation}
which is a keen probe of  the impact of explicit chiral symmetry breaking on the mass function. The Euclidean constituent-quark mass is obtained from the condition,
\begin{equation}
  (M^E)^2 :=  \{ s  | s = M^2(s) \} \ ,
\end{equation}
and depicted in Figure~\ref{flavourmassfunc} by the dashed (green) line intersecting with the mass-function solutions from which one finds (in GeV)~\cite{Roberts:2007jh}:
\begin{equation}
\begin{array}{c|ccccc}
  f & \mathrm{\ chiral} & u, d  & s  &  c &  b \\ \hline
  M_f^E  &  0.42 & 0.42 & 0.56 & 1.57 & 4.68 
\end{array}
\end{equation}

The ratio $\zeta_f $ thus quantifies the effect of explicit chiral symmetry breaking on the dressed- quark mass-function compared with the sum 
of the effects of explicit and dynamical chiral symmetry breaking. Calculation reveals $\zeta_d = 0.02$, $\zeta_s = 0.23$, $\zeta_c = 0.65$ and 
$\zeta_b = 0.8$, results which are readily understood. This constituent-sigma term vanishes in the chiral limit as expected,  is small for light quarks 
because the magnitude of their constituent-mass owes primarily to DCSB and approaches unity for heavy quarks because explicit chiral symmetry 
breaking becomes  the dominant source of their mass.

\section{Heavy mesons and Bethe-Salpeter amplitudes}
\label{sec3}

A meson appears as a pole in the four-point quark-antiquark Green function whose residue provides its Bethe-Salpeter amplitude (BSA). This amplitude 
can be determined reliably by solving the homogeneous renormalized Bethe-Salpeter equation (BSE) in a truncation scheme consistent with that employed 
in the gap equation~(\ref{DSEquark}). Particular forms of the BSE for the lightest pseudoscalar mesons and the important  Ward-Takahashi identities the 
solutions of the BSEs must satisfy are discussed in detail in Refs.~\cite{Maris:1997tm,Maris:2003vk}. We here summarize characteristic features of the 
BSA solutions. The pseudoscalar, $0^{-+}$, bound-state BSA can be decomposed into four Lorentz invariant amplitudes, 
\begin{equation}
      \Gamma_{H_{0^-}}(k;P) = \gamma_5 \left[ i E_{H_{0^-}}(k;P) + \gamma\cdot P F_{H_{0^-}}(k;P) + \gamma\cdot k \, G_{H_{0^-}}(k;P) - 
               \sigma_{\mu\nu} k_\mu P_\nu H_{H_{0^-}}(k;P) \right ] \ ,
\label{genGpi}
\end{equation}
where $P=p_1+p_2$ is the total-momentum entering the vertex and $k$ is the relative-momentum between the amputated quark legs. For bound states 
of constituents with equal current-quark masses the scalar functions $E, F, G$ and $H$ are even under $k\cdot P \to - k\cdot P$. In case of vector mesons 
the number of invariant amplitudes is 8 due to the spin content and the transversality of the vector-meson BSA~\cite{Maris:1999nt}. The axial-vector
Ward-Takahashi identity, which expresses chiral symmetry and its breaking pattern, implies an intimate relation between the kernel in the gap equation 
and that in the BSE. Indeed, it serves to prove a series of Goldberger-Treiman type of relations, one of which reads in the chiral limit for $H=\pi, K$:
\begin{equation}
  f_{H_{0^-}}^0 E_{H_{0^-}}(k;0)  \ =  \  B^0(k^2)
\label{GTrelation}
\end{equation}
Here, the superscript denotes the chiral limit of the quark propagator's scalar piece and of the weak decay constant, $f_{H_{0^-}}$, given by,
\begin{equation}
  f_{H_{0^-}} P_\mu = \langle 0 | \bar q_{f_2} \gamma_5 \gamma_\mu q_{f_1} |H_{0^-}\rangle  \ = \ Z_2\; \mathrm{tr}_\mathrm{CD}
 \int_q^\Lambda \! i\,\gamma_5\gamma_\mu \, S_{f_1}(q_+)\, \Gamma_{H_{0^-}}(q;P)\, S_{f_2}(q_-) \ 
\label{fpigen}
\end{equation}
where $q_+ =  q +\eta_P P$, $q_- = q- (1-\eta_P) P$, the trace is over Dirac and color indices and $f_i$ denotes the flavor content.

Of course, relation (\ref{GTrelation}) does not hold anymore for heavy-light mesons. Nonetheless, it is crucial to preserve the chiral properties of the light quark in
the determination of heavy-meson properties, because studies which fail to implement light-quark dressing consistently run into artifacts caused  by rather large 
constituent-quark masses and their associated propagator pole, $S(k)=(i\feynslash k + m_q)^{-1}$~\cite{ElBennich:2008xy,ElBennich:2008qa}. 
This, in turn, leads to considerable model dependence of heavy-light form factors and couplings~\cite{ElBennich:2009vx}.

The important asymmetry in quark masses of flavor-nonsinglet $Q\bar q$ mesons is at the origin of several diverse energy scales, a feature leading
to difficulties not encountered in the calculation of either light or heavy equal-mass DSEs and BSAs within the rainbow-ladder truncation. For example, while 
experimental mass values of flavored pseudoscalar mesons, such as $D_{(s)}$ and $B_{(s)}$ mesons, are easily reproduced, this is not the case for their 
respective weak decay constants~\cite{Maris:2005tt,Nguyen:2010yh}. It was already noted that the heavy quarks, and in particular the $b$-quark, have mass 
functions whose momentum dependence is nearly constant on a large domain. Thus, it is not unexpected that good agreement can be reached for the 
heavy-meson masses.

More precisely, within the context of the dressed propagators in Eq.~(\ref{sigmaSV}), it was shown~\cite{Maris:2005tt} that for small current-quark masses, 
weak decay constants of pseudoscalar and vector heavy-heavy and heavy-light mesons increase with the quark mass, yet tend to level off between the $s$- and 
$c$-quark mass. This behavior is consistent with the heavy-quark limit discussed in Section~\ref{sec1}; {\em i.e.}\/, in this limit the decay constant decreases  
with increasing meson mass like $f_{H}  \propto 1/\surd{m_H}$~\cite{Manohar:2000dt}.  This asymptotic behavior might occur as low as $Q=c$ for the 
$Q\bar u$ mesons. However, the decay constants $f_D$ and $f_{D_s}$ are about 20\% below their experimental values~\cite{Maris:2005tt}. It is also noteworthy
that $Q\bar q$ decay constants depend on the norm of the Bethe-Salpeter amplitudes and are thus proportional to the derivative of the quark propagators, 
it is not surprising that they are more sensitive to details of the model than the meson masses. They are thus better indicators for deficiencies in the modeling.

This strongly suggests that the rainbow-ladder truncation is not reliable and/or the model for the effective interaction in the $Q\bar q$ Bethe-Salpeter kernel  
is not applicable in the charm-quark region and even beyond,\footnote{\, A more detailed discussion about the failure of the rainbow-ladder truncation when 
dealing with $Q\bar q$ meson observables can be found in Section IX-D of Ref.~\cite{Bashir:2012fs}.} despite the fact that experimental meson masses are well 
reproduced. In fact, in such systems cancelations, which largely mask the effect of dressing the quark-gluon vertices, are blocked by the dressed-propagator 
asymmetry. The task to take the next step beyond the rainbow-ladder approximation may be facilitated by the exact form of the axial-vector Bethe-Salpeter 
equation,  valid when the quark-gluon vertex is fully dressed~\cite{Chang:2009zb}.

\section{Breaking of spin-flavor symmetries in heavy mesons}
\label{sec4}

The necessary ingredients to study heavy-meson phenomenology and calculate flavor-non-singlet observables, such as couplings and transition form factors 
between heavy and light(er) mesons, were presented in the two previous sections. We now turn our attention to these observables and scrutinize the validity 
of the heavy quark spin-flavor symmetries invoked earlier on in Section~\ref{sec1}.

In Section~\ref{sec3} we discussed the obstacles in defining an appropriate symmetry-preserving truncation scheme for the Bethe-Salpeter kernel of $Q\bar q$
mesons. This has so far impeded efforts to build adequate BSAs for heavy $0^{-+}$ and $1^{--}$ mesons. Calculations are for now limited to BSA models with 
a single width parameter commonly fixed by a fit to extant hadronic data. Nonetheless, the form factors calculated with the DSE model are obtained for the entire 
physical momentum domain without any extrapolations and the chiral limit is directly accessible~\cite{Ivanov:2007cw,ElBennich:2010ha,ElBennich:2011py,ElBennich:2009vx}. 
The simplest $Q\bar q$ observables that can be calculated are the weak $0^{-+}$ and $1^{--}$ meson decays constants. Although the rainbow-ladder truncation of 
the Bethe-Salpeter kernel is not successful in reproducing the existing experimental $0^{-+}$ decay constant values~\cite{Amsler:2008zzb}, the results in 
Refs.~\cite{Maris:2005tt,Nguyen:2010yh} as well as simulations in lattice-regularized QCD~\cite{Albertus:2010nm,Dimopoulos:2011gx,Becirevic:2012ti,Becirevic:1998ua}
clearly indicate that both heavy-quark spin and flavor symmetries are broken to a considerable extent in the charm and beauty sector. Consider for instance the ratios
of decay constants, 
\medskip

\begin{tabular}{clccl}
 $f_{D_s}/f_D =  1. 28$ &  (\textrm{DSE-RL model}~\cite{Nguyen:2010yh})  , & &  $f_{B_s}/f_B =  1. 37$ &  (\textrm{DSE-RL model}~\cite{Nguyen:2010yh})   ,\\
 $f_{D_s}/f_D =  1.17$  &  (\textrm{Lattice QCD}~\cite{Dimopoulos:2011gx})  , & &  $f_{B_s}/f_B =  1.15$  &  (\textrm{Lattice QCD}~\cite{Albertus:2010nm}) ,  \\
 $f_{D_s}/f_D =  1.19$  &  (\textrm{Lattice QCD}~\cite{Becirevic:2012ti}) , & & $f_{B_s}/f_B =  1.19$  &  (\textrm{Lattice QCD}~\cite{Dimopoulos:2011gx})  , \\
 $f_{D_s}/f_D =  1.32$  &  (\textrm{Experiment}~\cite{Amsler:2008zzb}) , &  & $f_{B_s}/f_B =  1.24$  &  (\textrm{NNLO ChPT}~\cite{Altenbuchinger:2011qn}) ,  \\
\end{tabular} 

\begin{tabular}{clccl}
 $f_{D^*}/f_D =  1. 04$  &  (\textrm{DSE-RL model}~\cite{Nguyen:2010yh}) , & &  $f_{B^*}/f_B =  1. 73$ &  (\textrm{DSE-RL model}~\cite{Nguyen:2010yh}) , \\
 $f_{D^*}/f_D =  1.28$  &  (\textrm{Lattice QCD}~\cite{Becirevic:2012ti})  , &  & $f_{B^*}/f_B =  1.11$  &  (\textrm{Lattice QCD}~\cite{Becirevic:1998ua}) , \\
\end{tabular}  \\

\noindent
where we omitted errors since they are only available in some cases. Even though there are considerable numerical differences among
the calculations and with experimentally extracted values, these ratios clearly illustrate that $SU(3)_F$ and spin-symmetry breaking effects are by no 
means insignificant.

Heavy-quark symmetry-breaking effects are also encountered in $b\to c$ and $c\to d$ transitions~\cite{Ivanov:2007cw} which occur in weak decays. 
They describe the nonperturbative matrix element between heavy and light(er) mesons which includes the propagation of the light spectator quark in the decays. 
Their precise evaluation is crucial in determining $CP$-violating observables in weak $D$ and $B$ decays~\cite{ElBennich:2006yi,Boito:2008zk,ElBennich:2009da} 
and oscillations~\cite{Leitner:2010fq,ElBennich:2011gm}. In semi-leptonic decays, the transition form factors fully describe the decay amplitude, 
{\em e.g.\/} in the case of heavy $H (0^{-+})$ to light(er) $P(0^{-+})$ transitions mediated by the weak HQET operators, $\bar q_l \gamma_\mu(1-\gamma_5) Q$, 
the matrix element is decomposed into two Lorentz vectors,
\begin{equation}
\langle  P(p_2) | \bar q_l \gamma_\mu (1 - \gamma_5 ) Q | H (p_1) \rangle \  =  \ F_+(q^2)  P_\mu   +   F_-(q^2) q_\mu  \ ,
\label{HPSformfac} 
\end{equation}
with the total heavy-meson momentum, $P_\mu=(p_1+p_2)_\mu$, $P^2=-M_H^2$, $q_\mu=(p_1-p_2)_\mu$, $Q=c,b$ and $q_l = u,d,s$. In the limit $m_Q\to \infty$, 
the mass function $M_Q(p^2)$ is constant, whereas for light quarks $M_q(p^2)$ is given by the solutions depicted in Figure~\ref{flavourmassfunc}. Thus, the use of a 
constituent-quark propagator with a constant dressed-mass function $\hat M_Q\simeq M^E_Q$ is a sensible approximation. If one assumes the validity of the expansion 
in Eq.~(\ref{mqpropagator}) and makes the two-covariant {\em Ansatz\/} limited to the pseudoscalar and pseudovector components of the heavy meson's BSA in Eq.~(\ref{genGpi}),
\begin{equation}
  \Gamma_{H_{0^-}} (k;P) = \gamma_5 \left [ 1 - \tfrac{1}{2}\, i\, \gamma\cdot \vv \right ]  \frac{1}{\mathcal{N}_H}\, \varphi (k^2) \  ,
\label{twocovariant}
\end{equation}
where $\mathcal{N}_H$ is the canonical normalisation constant and $P_\mu = (\hat M_Q +E )\vv_\mu$, $E = M_H - \hat M_Q$, in this limit, then the weak decay 
constant in Eq.~(\ref{fpigen}) becomes\footnote{\, The same expression holds for the vector meson decay constant at leading order in the $1/\hat M_Q$ expansion.
Analogously, the $H (0^{-+})$ to $V(1^{--})$ transition form factors are described by the same universal function $\xi (w)$ of Eq.~(\ref{formfactrans2}). }
with $z=u-2E\sqrt{u}$:
\begin{equation}
   f_{H_{0^-}}  = \frac{\kappa}{\sqrt{M_H}} \frac{N_c}{8\pi^2} \int_0^\infty \!\!du \left (\sqrt{u} -E \right ) \varphi^2(z) 
                            \left [ \sigma_S(z) + \tfrac{1}{2}\, \sqrt{u}\, \sigma_V(z) \right ] .
\end{equation}
We introduced the $M_H$-independent canonical normalization, $\kappa$, in order to make explicit the previously mentioned relation $f_{H_{0^-}}\sqrt{M_H} =$ constant.

Similarly, at leading order in $1/\hat M_Q$, the $H\to P$ transition form factors in Eq.~(\ref{HPSformfac}) are,
\begin{equation}
   F_{\pm} (q^2) \ = \ \frac{1}{2}\frac{M_P \pm M_H}{M_P M_H}\,  \xi (w) ,
 \label{formfactrans1}
\end{equation}
{\em i.e\/}. the form factors depend on a single universal function (the so-called Isgur-Wise function):
\begin{eqnarray}
   \xi (w) &  =  & \kappa^2  \frac{N_c}{32\pi^2} \int_0^1\!\! d\tau\ \frac{1}{W} \int_0^\infty \!\!\! du\, \varphi^2(z_w) 
   \left [ \sigma_S(z_w) + \sqrt{\frac{u}{W}} \, \sigma_V(z_w) \right ] \ ,
  \label{formfactrans2} 
\end{eqnarray}
with $W = 1+2\tau(1-\tau )(w-1)$,  $z_w = u-2E \sqrt{u/W}$ and $w = - \vv_H\cdot \vv_P = (M_H^2+M_P^2 - q^2)/(2M_H M_P)$. Owing to the canonical
normalization of the BSAs, $\xi(w=1) =1$. 

Thus, in the DSE modeling approach the results of heavy-quark symmetry are reproduced. Since the form factors are calculated both ways, namely with the full propagators 
and in the heavy-quark limit,  we are able to examine the fidelity of the formulae in Eqs.~(\ref{formfactrans1}) and (\ref{formfactrans2}) in this limit. Indeed, corrections to the 
heavy-quark symmetry limit of the order of $\lesssim 30$\% are encountered in $b\to c$ transitions and can be as large as a factor of 2 in $c\to d$ transitions, as verified in a 
vast array of light- and heavy-meson observables~\cite{Ivanov:2007cw}. Moreover, the following ratios of transition form factors  serve as a measure of $SU(3)_F$ breaking,
\begin{equation}
 \frac{F_+^{B\to K}(0)}{F_+^{B\to\pi}(0)} = 1.23 \ , \quad   \frac{A_0^{B\to K^*}(0)}{A_0^{B\to\rho}(0)} = 1.25 \ ,
\end{equation}
here taken at $q^2=0$ where the $A_0(q^2)$ are the appropriate form factors in $H (0^{-+})$ to $V(0^{--})$ transitions~\cite{Ivanov:2007cw}. The flavor breaking is
of similar order as in the decay constant ratios, $f_{D_s}/f_D$ and $f_{B_s}/f_B$, discussed above.

Flavored matrix elements also enter in the calculation of couplings between $D^{(\ast)}$- and light-pseudoscalar- and vector-mesons of effective heavy-meson 
Lagrangians. The models are typically an $SU(4)_F$ extension of light-flavor chirally-symmetric Lagrangians~\cite{Casalbuoni:1996pg}.  For instance, exotic states formed 
by heavy mesons and a nucleon were the object of a study based upon heavy-meson chiral perturbation theory~\cite{Yamaguchi:2011xb} in which a 
universal coupling, $\hat g$, between a heavy quark and a light pseudoscalar or vector meson was inferred from the strong decay $D^*\to D\pi$. 
The strong decays, $H^*(p_1) \to H(p_2) \pi(q),$ can generally be described by the invariant amplitude,
\begin{equation}
   \mathcal{A} (H^*\to H\pi)  \ =  \ \epsilon_\mu^{\lambda_{H^*}}\!(p_1)\ M_{\mu\nu}^{H^*H\pi}  \  :=  \  \epsilon_\mu^{\lambda_{H^*}}\!(p_1) \,  q_\mu \  g_{H^*H\pi} \ ,
\label{Hstarpicoupl}
\end{equation}
which at leading-order in a systematic, nonperturbative, symmetry-preserving DSE truncation scheme \cite{Munczek:1994zz,Bender:1996bb} is given by
\begin{equation}
  g_{H^*\!H\pi}\  \mathbf{\epsilon}^\lambda\!\!\cdot q   =   \mathrm{tr}_\mathrm{CD}\!  \int\! \frac{d^4k}{(2\pi)^4} \, \mathbf{\epsilon}^\lambda\!\cdot \Gamma_{H^*}(k;p_1)  S_Q(k+p_1)
            \bar  \Gamma_H(k;-p_2) S_f(k+q) \bar \Gamma_\pi(k;-q) S_f(k) \ ,
\label{eq2}
\end{equation}
where the trace is over color and Dirac indices, $q=p_1-p_2$, $\mathrm{\epsilon}^\lambda_\mu$ is the vector-meson polarization four-vector; 
and $S$ and $\Gamma$ are dressed-quark propagators and meson BSAs as described in Eqs.~(\ref{sigmaSV}) and (\ref{genGpi}), respectively.

The dimensionless coupling $g_{H^*H\pi}$, which can be determined via the decay width $\Gamma_{H^*H\pi}$, is related to the putative universal strong coupling
$\hat g$~\cite{Casalbuoni:1996pg}. At tree level, the couplings $g_{H^*\!H\pi}$ and $\hat g$ are related as,
\begin{equation}
    g_{H^*H\pi} \ = \  2\,\frac{\sqrt{m_H m_{H^*}}}{f_\pi}\, \hat g \, .
\end{equation}
Practically, the matrix element in Eq.\,\eqref{Hstarpicoupl} describes the physical process $D^*\to D\pi$, with both the final pseudoscalar mesons on-shell.  It also serves to 
compute the unphysical soft-pion emission amplitude $B^*\to B\pi$ in the chiral limit ($m_{B^*} - m_B < m_\pi$), which defines  $g_{B^*\!B\pi}$.  A comparison between these 
two couplings is an indication of the degree to which notions of heavy-quark symmetry can be applied in the charm sector. This was done in the impulse approximation using the 
simple heavy constituent-quark propagator mentioned above Eq.~(\ref{twocovariant}), which demonstrated that the difference in either extracting $\hat g$ from $g_{D^*\! D\pi}$ 
or $g_{B^*\!B\pi}$ is material~\cite{ElBennich:2010ha}:
\begin{equation}
   g_{D^*\!D\pi}   =15.8^{+2.1}_{-1.0}\ \Rightarrow \ \hat g = 0.53^{+0.07}_{-0.03}  \ ,   \quad  g_{B^*\!B\pi}  = 30.0^{+3.2}_{-1.4} \ \Rightarrow \ \hat g = 0.37^{+0.04}_{-0.02} \ .
\label{eqcouple1}    
\end{equation}
One can also employ a confining heavy-quark propagator of the kind,
\begin{equation}
\label{SQ}
  S_Q (k) \ =  \ \frac{-i \gamma\cdot k + \hat M_Q}{\hat M_Q^2}\  \mathcal{F}(k^2/\hat M_Q^2)\ , 
\end{equation}
with the definition ${\cal F}(x)= [1-\exp(-x)]/x$; this implements confinement but produces a momentum independent heavy-quark mass-function, {\em viz\/}. 
$\sigma_V^Q(k^2)/\sigma_S^Q(k^2)=\hat M_Q$. In this case we obtain with $\hat M_c = 1.32$~GeV and $\hat M_b = 4.65$~GeV  \cite{Ivanov:2007cw},
\begin{equation}
   g_{D^*\!D\pi}   =18.7^{+2.5}_{-1.4} \ \Rightarrow \ \hat g = 0.63^{+0.08}_{-0.05}  \ ,   \quad  g_{B^*\!B\pi}  = 31.8^{+4.1}_{-2.8} \ \Rightarrow \ \hat g = 0.39^{+0.05}_{-0.03} \ .
\label{eqcouple2}   
\end{equation}
From the experimental decay width one extracts: $g_{D^*\!D\pi}^\mathrm{exp.} = 17.9\pm1.9$ \cite{Anastassov:2001cw}. The couplings obtained with the confining propagators 
in Eq.~(\ref{eqcouple2}) are larger than those in Eq.~(\ref{eqcouple1}), yet both results emphasize that when the $c$-quark is a system's heaviest constituent, 
$\Lambda_\mathrm{QCD}/m_c$-corrections are not under good control.

\begin{figure}[t]
\vspace*{-2ex}
\centerline{\includegraphics[clip,width=0.47\textwidth]{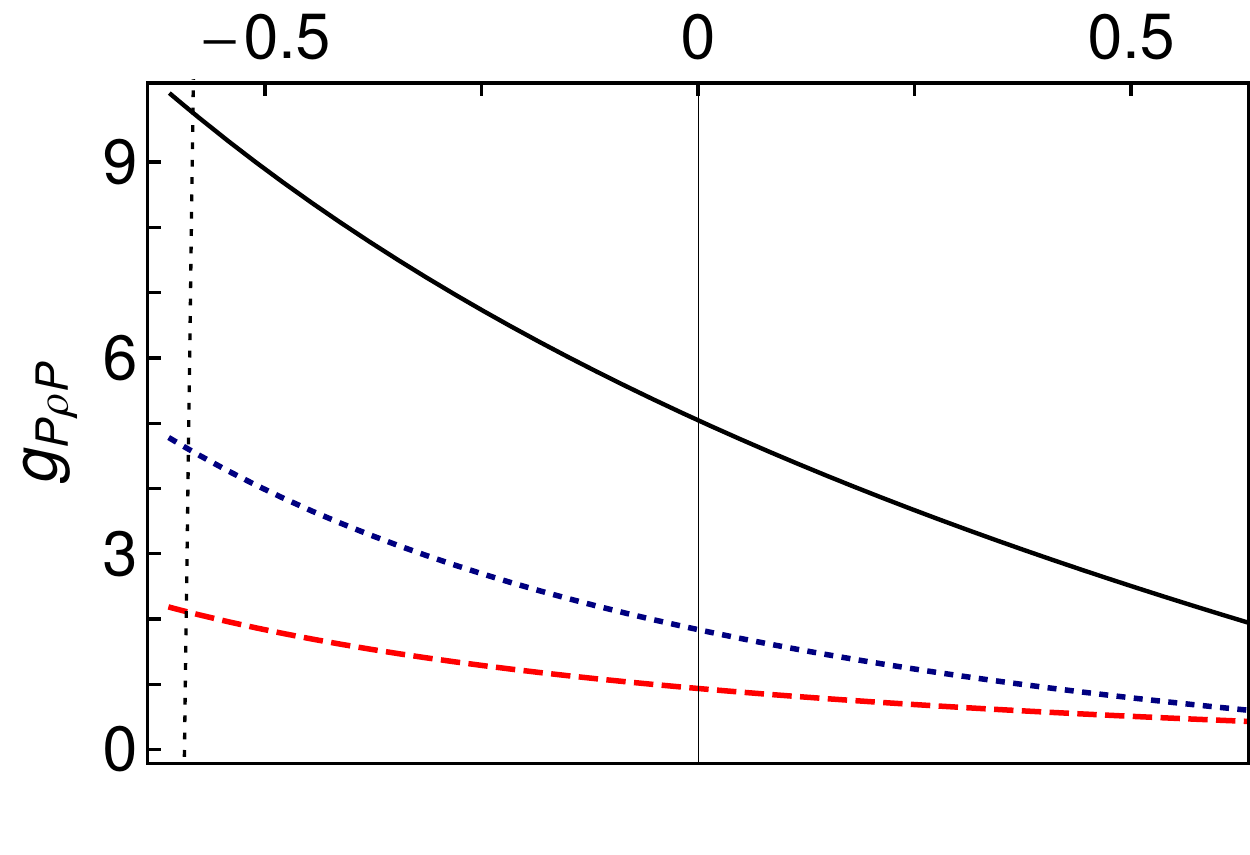}}
\vspace*{-3.3ex}

\centerline{\includegraphics[clip,width=0.473\textwidth]{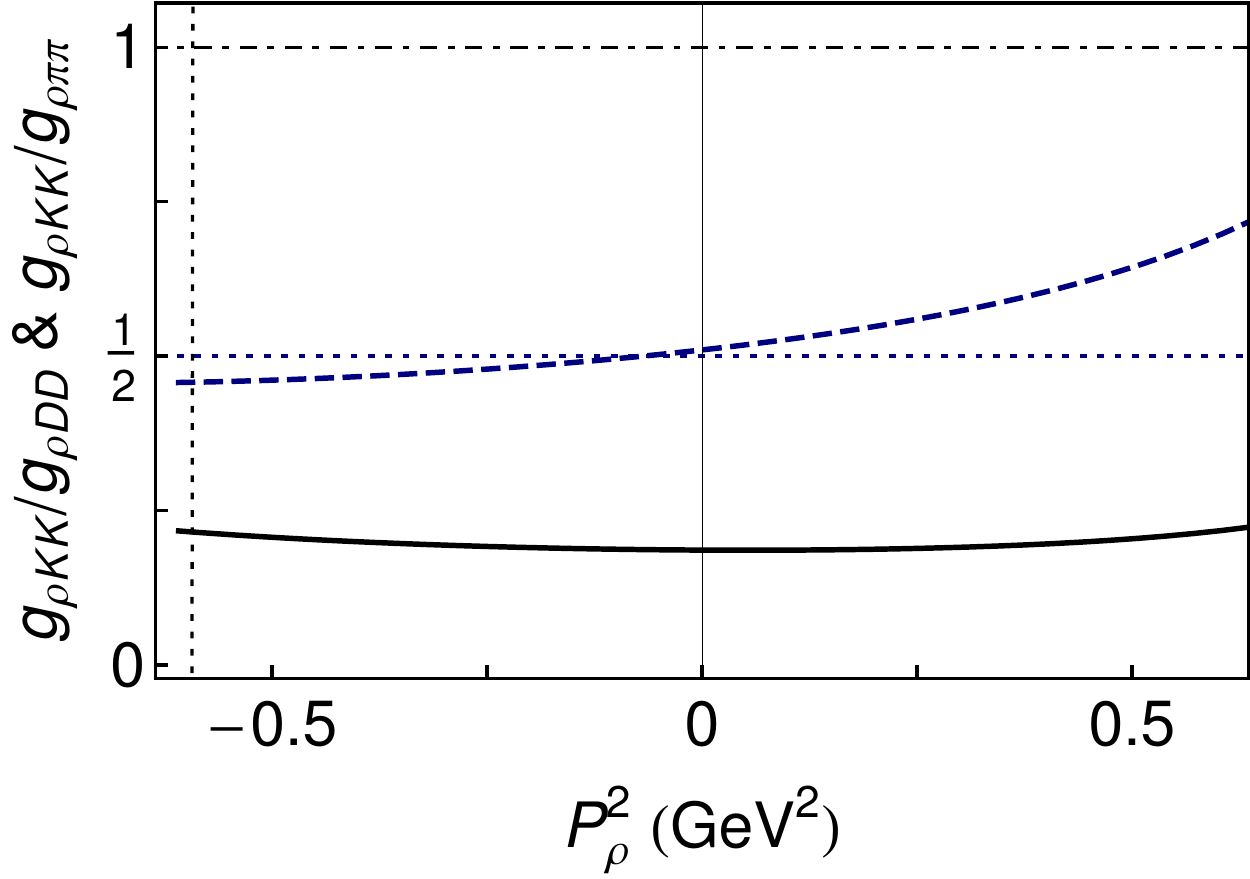}\hspace*{0.15em}}
\caption{
\emph{Upper panel} -- Dimensionless couplings: $g_{D\rho D}$ (solid curve); $g_{K\rho K}$ (dashed curve); and $g_{\pi\rho\pi}$ (dotted curve) -- all computed as a function 
of the $\rho$-meson's off-shell four-momentum-squared, with the pseudoscalar mesons on-shell.
\emph{Lower panel} -- Ratios of couplings: $g_{K\rho K}/g_{D\rho D}$ (solid curve); and $g_{K\rho K}/g_{\pi\rho \pi}$ (dashed curve).  In the case of exact $SU(4)$ symmetry, 
these ratios take the values, respectively, $1$ (dot-dashed line) and $(1/2)$ (dotted line). The vertical dotted line marks the $\rho$-meson's on-shell point in both panels.}
\label{figcoupling}
\end{figure}

We here take the opportunity to extend this study to the two unphysical strong decay amplitudes $B^*_s\to BK$ and $D^*_s\to DK$, from which one can extract the effective 
couplings $g_{D_s^*DK}$ and $g_{B_s^*BK}$, where the latter was recently calculated with QCD sum rules~\cite{Cerqueira:2011za}. We remind that in Eq.~(\ref{eq2}) the 
amplitude can be calculated for any value of $q^2$ and the chiral limit is directly accessible --- we do not need to resort to extrapolations, neither from spacelike to timelike 
momenta nor in current-quark mass, expedients which are necessary in other approaches~\cite{Becirevic:2009xp,Cerqueira:2011za,Bracco:2011pg}. Our results are:
\begin{equation}
   g_{D_s^*DK}   = 24.1^{+2.5}_{-1.6}   \ ,   \quad  g_{B_s^*BK}  = 33.3_{-3.7}^{+4.0} \  .
\label{eqcouple3}   
\end{equation}
Evaluating the coupling ratios, $g_{D_s^*DK} / g_{D^*\!D\pi}$ and $g_{B_s^*BK} / g_{B^*\!B\pi}$, with the values in Eq.~(\ref{eqcouple2}) and Eq.~(\ref{eqcouple3}), 
the order of magnitude of $SU(3)_F$ breaking is found to be similar to that of the decay constant ratios $f_{D_s}/f_D$ and $f_{B_s}/f_B$. It is notable that our result 
for $g_{B_s^*BK}$ is perfectly consistent with that for $g_{B^*\!B\pi}$ in Eq.~(\ref{eqcouple2}) but stands in stark contrast to the value of Ref.~\cite{Cerqueira:2011za}: 
$g_{B_s^*\!BK} = 10.6 \pm 1.7$. 

To conclude this section, we turn our attention to a direct measure of $SU(4)_F$-symmetry, the accuracy of which can be checked in relations between $\pi \rho \pi$, 
$K \rho K$ and $D\rho D$ couplings~\cite{ElBennich:2011py}. Clearly, having appreciated the significance of flavor $SU(3)_F$ breaking, we expect $SU(4)_F$ to be
even more strongly violated~\cite{Bracco:2011pg}. Indeed, as highlighted in Figure~\ref{figcoupling}, in the case of exact $SU(3)_F$ symmetry one would have 
$g_{K \rho K} = g_{\pi\rho \pi}/2$. This assumption provides a fair approximation on a domain $P^2\in [-m_\rho^2,m_\rho^2]$ where the deviation ranges from $(-10)\,$--$40\,$\%. 
Moreover, if $SU(4)_F$ symmetry were exact, then $g_{D\rho D} = g_{K \rho K} = g_{\pi\rho \pi}/2$, but in Ref.~\cite{ElBennich:2011py} the expectation $g_{D\rho D} = g_{K \rho K}$ 
is found to be violated on the entire domain $P^2\in [-m_\rho^2,m_\rho^2]$ at a level of between $360\,$--$\,440$\%.  The second identity, $g_{D\rho D}=g_{\pi\rho \pi}/2$, 
is violated at the level of $320\,$--$\,540$\%.

\section{Closing remarks}

Symmetries are a dominant, recurrent theme in physics, and it is for good reason that they have been such a longstanding focus of attention. Nonetheless, a given symmetry may 
not always be explicit in the appropriate Lagrangian. We have discussed two kinds of symmetries and the breaking thereof which are not apparent in the QCD Lagrangian: heavy-quark
spin and flavor symmetries emerge only once the heavy sector of QCD is expanded in terms of inverse powers of the heavy-quark mass, where next-to-leading orders can already 
contribute significantly to their breaking; and whereas in massless QCD chiral symmetry is a property of the Lagrangian, DCSB is an emergent phenomenon with capital consequences
for the theory: it lies at the origin of a mass scale in QCD, gives the Goldstone mode its mass and is responsible for most of the matter we encounter in nature. In reviewing
heavy-quark symmetries and the nonperturbative DSE approach to their qualitatively and quantitatively observable properties, we conclude that HQET is a poor guide to 
charm physics. The charm quark is, as has long been known, neither a light nor a heavy enough quark. With respect to the $b$-quark, while HQET is a sensible tool, we insist 
that reliable calculations of heavy-light $b\bar q$ form factors, couplings and decay constants require the veracious description of DCSB in the light quark's propagation.

\acknowledgments{We appreciated helpful discussions with Gast\~ao Krein. B.~E. acknowledges support from FAPESP, grant nos.~2009/51296-1 and 2010/05772-3. M.~A.~I.  
appreciates  the partial support of the Russian Fund of Basic Research, grant no.~10-02-00368-a. C.~D.~R. is supported by U. S. Department of Energy, Office of Nuclear Physics, 
contract no.~DE-AC02-06CH11357.

\end{document}